\documentclass[10pt,NJP]{iopart}

\usepackage{graphicx}
\usepackage{graphicx,color}

\expandafter\let\csname equation*\endcsname\relax
  \expandafter\let\csname endequation*\endcsname\relax
\usepackage{amsmath}
\usepackage{color}

\usepackage{amssymb}
\usepackage{amsfonts}
\usepackage{bm}
\usepackage[T1]{fontenc}
\usepackage[utf8]{inputenc}
\usepackage{iopams}
\def\<{\langle}
\def\>{\rangle}

\newcommand{\tens}[1]{\mbox{\textbf{\textit{\textsf{#1}}}}}

\definecolor{darkgreen}{rgb}{0.27451, 0.458824, 0.22745}
\usepackage{soul}
\usepackage{color}

\bibstyle{unsrt}

\begin{document}

\setlength{\parindent}{0cm}

\title{Manipulating the Coulomb interaction: A Green's function perspective}

\author{Pablo Barcellona, Robert Bennett}

\address{Physikalisches Institut, Albert-Ludwigs-Universit\"at
Freiburg, Hermann-Herder-Str. 3, 79104 Freiburg, Germany}

\author{Stefan Yoshi Buhmann}

\address{Physikalisches Institut, Albert-Ludwigs-Universit\"at
Freiburg, Hermann-Herder-Str. 3, 79104 Freiburg, Germany}

\address{Freiburg Institute for Advanced Studies,
Albert-Ludwigs-Universit\"at Freiburg, Albertstr. 19, 79104 Freiburg,
Germany}

\begin{abstract}
We present a unified framework for studying Coulomb interactions in arbitrary environments using macroscopic quantum electrodynamics on the basis of the electromagnetic
Green's function.
Our theory can be used to derive the Coulomb potential of a single charged particle as well as that between two charges in the presence
of media, bodies and interfaces of arbitrary shapes. To demonstrate this, we reproduce the well-known screened Coulomb force, account for local-field effects
and consider new cases such as a dielectric cavity and a conducting plate with a hole.

\end{abstract}
\submitto{\NJP}

\maketitle

\section{Introduction}

The Coulomb force is perhaps the first electromagnetic interaction encountered by a student of physics, introduced as an immutable inverse square law that delivers the force between two charged particles. The simple inverse square distance dependence is used as the basis for a wide variety of descriptions of nature, from the Hartree-Fock methods of quantum chemistry to the Derjaguin, Landau, Verwey, and Overbeek (DLVO) theory of colloidal dispersions \cite{Derjaguin,Verwey}. What is usually missing is the fact that no physical system exists in true isolation; there will always be some environment enclosing the objects of interest. A prominent example is the exponential screening of the Coulomb interaction for charges embedded in a non-local medium \cite{tarasov}. This appears across physics as such effects arise in materials with a high density of free or quasi-free charge carriers; these include metals (where the screening length is the Thomas-Fermi length), electrolytes (where the screening length is the Debye-Huckel length), ionic solutions or narrow-band-gap semiconductors \cite{Thomas,Fermi,Fermi2,ashcroft}. One common way of arriving at these effects is to make a Thomas-Fermi approximation for a free electron gas, then solving the resulting screened Poisson equation. A small number of works in colloid physics \cite{Hurd} exist that extend this to charges near an interface, but these rely on a linearised Poisson-Boltzmann equation in which several assumptions must be made. \\

A more fundamental and flexible point of view is provided by macroscopic quantum electrodynamics (QED), where the Coulomb interaction involves the emission and reabsorption of a photon. For example, the interaction between two charges is mediated by a photon that is emitted from one charge and subsequently absorbed by the other. On the way, this photon may interact with its environment, for example it may reflect off a macroscopic body or be travelling through some bulk media. This leads to an environment-dependent Coulomb force, which is the subject of this work. We will write the Coulomb force in terms of a version of the dyadic Green's function, well-known from the formalism known as macroscopic QED \cite{Buhmann,gruner}. The special case of bulk media will, as we shall see, reproduce the screening effects discussed above, but the unified formalism we use is much more general. It provides a link between medium-assisted Coulomb interactions and the considerable literature on dyadic Green's functions \cite{tai,chew}. We will demonstrate this by considering general expressions as well as several particular geometries that demonstrate the power of the toolbox we are presenting. \\

   In general, environment-dependent effects may be described by the dielectric function, so if this can be engineered the Coulomb interaction can be controlled. One simple way to do this is by varying macroscopic geometric parameters \cite{karakasoglu}, but an increasingly relevant class of media are those whose microscopic structure is designed to generate a desired permittivity. Control of the Coulomb interaction is particularly important in many solid-state devices like solar cells, where an increase of screening allows for a stronger separation of excitons into electrons and holes \cite{gregg}. 
Moreover trapped atomic ions can be used as a collection of qubits where the quantum information is transferred between the ions thorough their mutual Coulomb interaction \cite{Mehta,Harty}.\\

An interesting related problem, which is of particular interest in colloid physics, is the Coulomb interaction between two charged particles near non translational-invariant media, like for example 
a planar multilayer system. This is distinct from the screening imposed by bulk media, and is much less well-studied, though in the formalism we present here it may be studied in exactly the same way as screening effects.
Charged particles near macroscopic objects induce a polarization-charge density on the surface of those objects, which in turn affects their Coulomb interaction. These polarization effects are particularly important at the interfaces between media with very different dielectric permittivities. This problem is usually treated by the method of images, where the medium is replaced by a set of image charges in order to satisfy a relevant set of boundary conditions on the surfaces. Sometimes, however, in complex geometries (such as, e.g. a wedge)  it is not clear where to place the image charges so one must resort to complex calculations of potentials for particular systems, which may be of limited applicability. Here we will express all Coulomb forces in terms of the dyadic Green's function, which is a very well-studied object in a large number of different geometries.
The interaction between two atoms near the relatively simple system of an infinite dielectric slab or metal has been studied \cite{allen,edmonds,efrima,apell} and, in the metallic case, extended to include spatial dispersion. The result shows that the Coulomb interaction must be corrected for distances shorter in comparison with the Thomas-Fermi screening length in the dispersive case \cite{gabovich}. Similarly, the Coulomb interaction between a charged moving particle and a plasma has been studied in the literature \cite{barton}. \\

Although some disparate particular cases have been investigated, to our knowledge no general expression of the medium-assisted Coulomb interaction is known in the literature. As stated earlier, the aim of this work is to study the Coulomb interaction in generic environments that are described by the dyadic Green's tensor. The interaction will be described in the framework of macroscopic QED as a one-photon exchange process, where the photon propagator is governed by the electromagnetic environment.\\

After deriving some general formulae, we will demonstrate the power of the method by firstly considering a set of examples that reproduce well-known results, the novelty arising from their unification within the same framework. These will include the screening for spatially dispersive media, the interaction between two charges near a planar interface between two dielectrics and the interaction between a charge and a polarizable particle.
Following this we also apply our general results to new, non trivial environments, namely local-field effects, a dielectric cavity and a plate with a hole. Our approach can be applied to every case where the Green tensor is known or can be calculated by analytical or numerical means.

  \section{Coulomb interaction in the presence of dielectrics}
  To study the Coulomb interaction between two charged particles, we use field quantization in linear absorbing and non-local media \cite{Buhmann,Buhmann2,buhnonloc,butcher}, using the Coulomb gauge. Here and throughout we take the medium to be non-magnetisable (i.e. with unit relative permeability), so that the electric field can be expanded in terms of the creation and annihilation operators $\hat{\mathbf{f}}^\dag\left( \mathbf{r},\omega  \right)$, $\hat{\mathbf{f}}\left( \mathbf{r},\omega  \right)$ for electric excitations labelled by frequency $\omega$ and position $\mathbf{r}$;
   \begin{equation}
\hat {\mathbf{E}}\left( \mathbf{r},\omega  \right) = \int \text{d}^3 s \tens{G}_e\left( \mathbf{r},\mathbf{s},\omega  \right) \cdot \hat{ \mathbf{f}}\left( \mathbf{s},\omega  \right)\; . 
   \end{equation}
The tensors $ \tens{G}_e$ are mode-tensors that depend on the imaginary parts of the electric permittivity of the absorbing medium:
\begin{equation}
\tens{G}_e\left( \mathbf{r},\mathbf{r}',\omega  \right) = \text{i}\frac{\omega ^2}{c^2}\sqrt {\frac{\hbar }{\pi \epsilon _0}\text{Im} \varepsilon \left( \mathbf{r}',\omega  \right)} \tens{G}\left( \mathbf{r},\mathbf{r}',\omega  \right)
\label{mode}
\end{equation}
and on $\tens{G}\left( \mathbf{r},\mathbf{r}',\omega  \right)$ the classical Green tensor  of the Helmholtz equation \cite{Buhmann,Buhmann2}:
\begin{equation}\label{FullG}
\left[ \nabla  \times \nabla  \times  - \frac{\omega ^2}{c^2}\epsilon \left( \mathbf{r},\omega  \right) \right] \cdot \tens{G}\left( \mathbf{r},\mathbf{r}',\omega  \right) = \delta \left( \mathbf{r} - \mathbf{r}' \right)\tens{I}
\end{equation}

An important relation involving these mode-tensors is:
  \begin{equation}
 \int \text{d}^3s \tens{G}_e\left( \mathbf{r},\mathbf{s},\omega  \right) \cdot \tens{G}_e ^{*\top} \left(\mathbf{r}', \mathbf{s},\omega  \right)  =\frac{\hbar \mu _0}{\pi }\omega ^2 \text{Im} \tens{G}\left( \mathbf{r},\mathbf{r}',\omega  \right),
\label{int-rel}
\end{equation} 
where $^{\top}$ denotes the transpose.

Charged particles interact with the radiation field through the scalar and vector potentials, not directly via the electric field. 
The scalar potential $\hat{\phi}$ is related to the longitudinal part of the electric field via $\hat{\mathbf{E}}^\parallel  =  - \nabla \hat \phi $. The longitudinal part $\mathbf{f}^\parallel $ of a general vector-valued function $\mathbf{f}$ can be calculated using the longitudinal delta function $\bm{\delta} ^\parallel (\textbf{r})=-\nabla \nabla (4 \pi |\mathbf{r}|)^{-1}$ via:
\begin{equation}
\mathbf{f}^\parallel ( \mathbf{r} ) = \int \text{d}^3 s \bm{\delta} ^\parallel ( \mathbf{r} - \mathbf{s}  ) \cdot \mathbf{f}\left( \mathbf{s} \right)
\end{equation}
Hence the scalar potential satisfies the equation:
\begin{equation}
\nabla \hat \phi\left( \mathbf{r},\omega  \right) =- \int \text{d}^3 s {^\parallel \tens{G}_e}\left( \mathbf{r},\mathbf{s},\omega  \right) \cdot \hat{ \mathbf{f}}\left( \mathbf{s},\omega  \right)
   \end{equation}
   where $^\parallel \tens{G}_e$ is the left-longitudinal component of $\tens{G}_e$:
   \begin{equation}
   ^\parallel \tens{G}_e \left( \mathbf{r},\mathbf{r}',\omega  \right) = \int \text{d}^3s \bm{\delta} ^\parallel ( \mathbf{r} - \mathbf{s}) \cdot \tens{G}_e\left( \mathbf{s},\mathbf{r}',\omega  \right)
   \end{equation}
  The scalar potential can hence be derived performing a line integral of a vector field:
\begin{equation}
\hat \phi \left( \mathbf{r},\omega  \right) =  - \int \text{d}^3 s\int\limits_{\mathbf{r}_0}^{\mathbf{r}} \text{d}\mathbf{r}'  \cdot {^\parallel \tens{G}}_e \left( \mathbf{r}',\mathbf{s},\omega  \right) \cdot \hat {\mathbf{f}}\left( \mathbf{s},\omega  \right)
\end{equation}
We assume that  the point $\textbf{r}_0$ may be placed at infinity ($|\textbf{r}-\textbf{r}_0|\to \infty$) and take the scalar potential to be defined relative to this; $\hat \phi \left( \mathbf{r}_0,\omega  \right)=0$. 
As expected, the potential at a point $\textbf{r}$ is proportional to the work done by the longitudinal electric field in order to move the  charge from infinity to that point.
Analogously the vector potential can be expressed in terms of the bosonic excitation operators:
\begin{equation}
\mathbf{\hat A}\left( \mathbf{r},\omega  \right)= \frac{1}{\text{i} \omega}  \int \text{d}^3 s {^\bot \tens{G}}_e \left( \mathbf{r},\mathbf{s},\omega  \right) \cdot \hat {\mathbf{f}}\left( \mathbf{s},\omega  \right)
\end{equation}
where $^\bot \tens{G}_e$ is the left-transverse component of $\tens{G}_e$:
   \begin{equation}\label{leftTransverseG}
   ^\bot \tens{G}_e \left( \mathbf{r},\mathbf{r}',\omega  \right) = \int \text{d}^3s \bm{\delta} ^\bot ( \mathbf{r} - \mathbf{s}) \cdot \tens{G}_e\left( \mathbf{s},\mathbf{r}',\omega  \right)
   \end{equation} 
   and $\bm{\delta} ^\bot (\textbf{r})=\nabla \times (\nabla \times \bm{I}) (4 \pi |\mathbf{r}|)^{-1}$ is the transverse delta function.

 \section{Energy shift}
 We consider the interaction between two charged particles assisted by a polarizable medium. 
 The distances between the particle and any interfaces of the medium are considered to be large enough that the interface may be regarded as a macroscopic surface, i.e. its microscopic structure is not resolves. This means that we are far enough away to exclude such effects as Pauli repulsion and covalent bonding. \\

\begin{figure}[ht]
   \centering
   \includegraphics[scale=1]{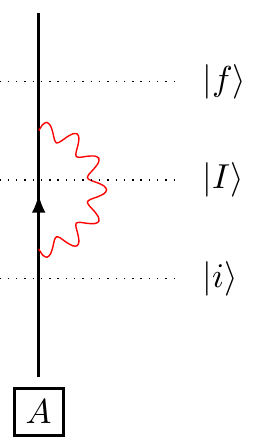}
   \caption{Single-atom Coulomb interaction.}
    \label{fig1}
   \end{figure}

As mentioned in the introduction, the Coulomb interaction may be pictured as arising from the emission and absorption of one virtual photon. There are four processes which can contribute to the interaction, which can be conveniently represented through Feynman diagrams. Two diagrams involve the photon being emitted and reabsorbed by the same particle as shown in see Fig \ref{fig1}. This photon may be affected by the medium, causing a position-dependent shift --- hence this process describes the interaction between one charged particle and a polarizable medium.

\begin{figure}[ht]
   \centering
   \includegraphics[scale=1.0]{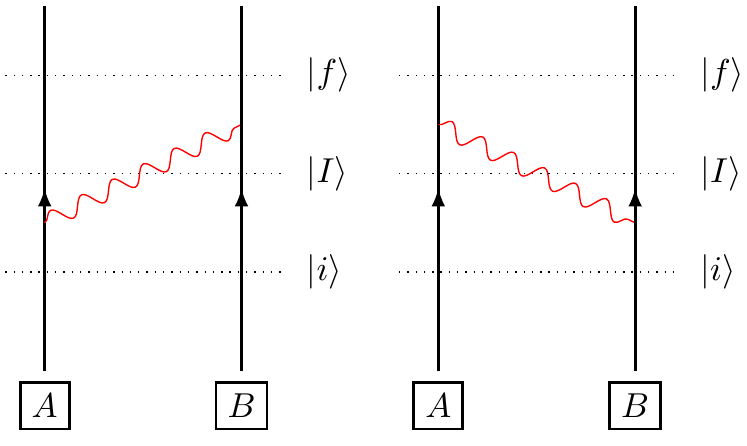}
   \caption{Medium-assisted Coulomb interaction between two atoms.}
    \label{fig2}
   \end{figure}

The other two diagrams involve the exchange of a single virtual photon between the two charges (see Fig. \ref{fig2}), and describe the Coulomb interaction between the pair. In all four of these diagrams the interaction is affected by the medium because the photon can be reflected by the body's surface, and hence can be considered emitted by a fictitious image charge.
%

The energy shift describing the Coulomb interaction can be obtained from second order perturbation theory 
\begin{equation}
\label{shift}
\Delta E_{(2)} =\langle i|\hat H_{\text{int}}|i\rangle  -\sum_{I\neq i}\frac{\langle i|\hat{H}_\textrm{int}|I\rangle\langle I|\hat{H}_\textrm{int}|i\rangle}{E_I-E_0},
\end{equation} 
and $\hat H_{\text{int}}$ is the Hamiltonian for a set of charges $q_i$ of momentum $\mathbf{p}_i$ minimally coupled to the electromagnetic scalar and vector potentials $\hat{\phi}$ and $\hat{\mathbf{A}}$;
\begin{equation}\label{Hint}
\hat H_{\text{int}} = \sum\limits_{i = A,B} q_i\hat \phi \left( \mathbf{r}_i \right)  - \sum\limits_{i = A,B} \frac{q_i}{m_i}\mathbf{\hat p}_i \cdot \mathbf{\hat A}\left( \mathbf{r}_i \right)  + \sum\limits_{i = A,B} \frac{q_i^2}{2m_i}\mathbf{\hat A}^2\left( \mathbf{r}_i \right) \; . 
\end{equation}

The second term is absent when one considers charges fixed in space ($\mathbf{ p}_i=0$).
  The intermediate state $|I\rangle$ corresponds to a state in which one virtual photon is present: $|I\rangle \!\equiv\! |\{\textbf{1}\left( \mathbf{r},\omega  \right)\}\rangle$, $E_I-E_0=\hbar \omega$, while the initial state corresponds to a state without photons $|i\rangle \!\equiv\! |\{0\}\rangle$.

Using the integral relation \eqref{int-rel}, the matrix elements of the individual terms that make up Eq.~\eqref{shift} read;
 \begin{align}
\langle i| \sum\limits_{i = A,B} q_i\hat \phi \left( \mathbf{r}_i \right)|I\rangle  =&  - \sum\limits_{i = A,B} q_i \int\limits_{\mathbf{r}_0}^{\mathbf{r}_i} \text{d} \mathbf{r}'  \cdot {^\parallel \tens{G}_e}\left( \mathbf{r}',\mathbf{r},\omega  \right)\\ 
\langle i|\sum\limits_{i = A,B} \frac{q_i^2}{2m_i}\mathbf{ \hat A}^2\left( \mathbf{r}_i \right) |i\rangle  =&
 \sum\limits_{i = A,B} \frac{\hbar \mu _0 q_i^2}{2\pi m_i}\int\limits_0^\infty  \text{d}\omega \text{Tr}\left\{ \text{Im} ^ \bot \tens{G}\left( \mathbf{r}_i,\mathbf{r}_i,\omega  \right)^ \bot  \right\}  
 \end{align}
 where we have introduced the two-sided-transverse component of $\tens{G}$
\begin{equation}
   ^\bot \tens{G}\left( \mathbf{r},\mathbf{r}',\omega  \right)^\bot = \int \text{d}^3s  \int \text{d}^3s'
 \bm{\delta} ^\bot \left( \mathbf{r} - \mathbf{s}\right) \cdot \tens{G}\left( \mathbf{s},\mathbf{s}',\omega  \right) \cdot \bm{\delta} ^\bot \left( \mathbf{s}' - \mathbf{r}'\right).
   \end{equation}
where $\bm{\delta} ^\bot$ is the transverse delta function defined below Eq.~\eqref{leftTransverseG}.
Using this and the integral relation \eqref{int-rel}, the total energy shift reads:
\begin{multline}
\label{ShiftG}
\Delta E_{(2)} =  - \frac{\mu _0}{\pi }\sum\limits_{i,j = A,B} q_i q_j  \int\limits_0^\infty  \frac{ \text{d}\omega}{\omega} 
 \omega ^2\int\limits_{\mathbf{r}_0}^{\mathbf{r}_i}  \int\limits_{\mathbf{r}_0}^{\mathbf{r}_j} \text{d}\mathbf{r}  \cdot \text{Im} ^\parallel \tens{G}\left( \mathbf{r},\mathbf{r}',\omega  \right)^\parallel  \cdot \text{d}\mathbf{r}' \\
 + \sum\limits_{i = A,B} \frac{\hbar \mu _0 q_i^2}{2\pi m_i}\int\limits_0^\infty  \text{d}\omega \text{Tr}\left\{ \text{Im} ^ \bot \tens{G}\left( \mathbf{r}_i,\mathbf{r}_i,\omega  \right)^ \bot  \right\}  
\end{multline}
where we have introduced the two-sided-longitudinal component of $\tens{G}$ :
\begin{equation}
   ^\parallel \tens{G}\left( \mathbf{r},\mathbf{r}',\omega  \right)^\parallel = \int \text{d}^3s  \int \text{d}^3s'
 \bm{\delta} ^\parallel \left( \mathbf{r} - \mathbf{s}\right) \cdot \tens{G}\left( \mathbf{s},\mathbf{s}',\omega  \right) \cdot \bm{\delta} ^\parallel \left( \mathbf{s}' - \mathbf{r}'\right).
   \end{equation}
where $\bm{\delta} ^\parallel=\nabla \otimes \nabla (4 \pi |\mathbf{r}|)^{-1}$ is the longitudinal delta function. 
We use the Schwarz reflection principle;
\begin{equation}
\tens{G}^*\left( \mathbf{r},\mathbf{r}',\omega  \right)=\tens{G}\left( \mathbf{r},\mathbf{r}',-\omega  \right)
\end{equation}
%
to extend the frequency integral \eqref{ShiftG} to the whole real axis :
\begin{multline}
\Delta E_{(2)} =  - \frac{\mu _0}{2\pi \text{i} }\sum\limits_{i,j = A,B} q_i q_j  \int\limits_{-\infty}^\infty  \frac{ \text{d}\omega}{\omega} 
 \omega ^2 \int\limits_{\mathbf{r}_0}^{\mathbf{r}_j}   \int\limits_{\mathbf{r}_0}^{\mathbf{r}_i} \text{d}\mathbf{r}  \cdot ^\parallel \tens{G}\left( \mathbf{r},\mathbf{r}',\omega  \right)^\parallel \cdot \text{d}\mathbf{r}'\\
  + \sum\limits_{i = A,B} \frac{\hbar \mu _0 q_i^2}{2\pi m_i}\int\limits_0^\infty  \text{d}\omega \text{Tr}\left\{ \text{Im} ^ \bot \tens{G}\left( \mathbf{r}_i,\mathbf{r}_i,\omega  \right)^ \bot  \right\}  
\end{multline}
The tensor $\omega^2\tens{G}\left( \mathbf{r},\mathbf{r}',\omega  \right)$ is analytic in the upper half of the
complex plane (including the real axis), and it is also finite at the origin. This means we close the path with an infinitely large half-circle in the upper complex half-plane and
deal with the singular point $\omega=0$ using the principal value prescription. The integral along the infinite semi-circle vanishes for $\textbf{r} \ne \textbf{r}'$ because:
\begin{equation}\mathop {\lim }\limits_{\left| \omega  \right| \to  
\infty }
\omega ^2\left. \tens{G}\left( \mathbf{r},\mathbf{r}',\omega  \right)
\right|_{\mathbf{r} \ne \mathbf{r}'} = 0\end{equation}
Calculating the residue at $\omega=0$, the energy shift reads:
\begin{multline}
\Delta E_{(2)} =  - \frac{1}{2 \epsilon_0}\sum\limits_{i,j = A,B} q_i q_j \int\limits_{\mathbf{r}_0}^{\mathbf{r}_i}  \int\limits_{\mathbf{r}_0}^{\mathbf{r}_j} \text{d}\mathbf{r}  \cdot \tens{G}\left( \mathbf{r},\mathbf{r}'  \right)\cdot  \text{d}\mathbf{r}' \\
 + \sum\limits_{i = A,B} \frac{\hbar \mu _0 q_i^2}{2\pi m_i}\int\limits_0^\infty  \text{d}\omega \text{Tr}\left\{ \text{Im} ^ \bot \tens{G}\left( \mathbf{r}_i,\mathbf{r}_i,\omega  \right)^ \bot  \right\} 
\label{energyshift}
\end{multline}
where we have defined the static Green tensor
\begin{equation}
\tens{G}\left( \mathbf{r},\mathbf{r}'  \right)= \mathop {\lim }\limits_{\omega  \to 0}  \frac{\omega ^2}{c^2} {^\parallel \tens{G}}\left( \mathbf{r},\mathbf{r}',\omega  \right)^\parallel 
=\mathop {\lim } \limits_{\omega  \to 0}\frac{\omega ^2}{c^2} \tens{G}\left( \mathbf{r},\mathbf{r}',\omega \right).
\label{g}
\end{equation}
The last equality follows from the fact that for zero frequency the Green's tensor is purely longitudinal \cite{Buhmann}.

The energy shift \eqref{energyshift} consists of two parts. Firstly there are the single-particle terms which describe the interaction with the surface;
\begin{multline}
U\left( \mathbf{r}_\text{A}\right) =  - \frac{q_\text{A}^2}{2 \epsilon_0}  \int\limits_{\mathbf{r}_0}^{\mathbf{r}_\text{A}} \int\limits_{\mathbf{r}_0}^{\mathbf{r}_\text{A}} \text{d}\mathbf{r} \cdot \tens{G}\left( \mathbf{r},\mathbf{r}'  \right) \cdot \text{d}\mathbf{r}' 
 + \frac{\hbar \mu _0 q_A^2}{2\pi m_A}\int\limits_0^\infty  \text{d}\omega \text{Tr}\left\{ \text{Im} ^ \bot \tens{G}\left( \mathbf{r}_A,\mathbf{r}_A,\omega  \right)^ \bot  \right\} 
\label{one-body}
\end{multline}
 and secondly the medium-assisted Coulomb interaction between the two particles, which reads:
\begin{equation}
U\left( \mathbf{r}_\text{A},\mathbf{r}_\text{B} \right) =  - \frac{q_\text{A} q_\text{B}}{ \epsilon_0} \int\limits_{\mathbf{r}_0}^{\mathbf{r}_\text{A}} \int\limits_{\mathbf{r}_0}^{\mathbf{r}_\text{B}} \text{d}\mathbf{r}  \cdot \tens{G}\left( \mathbf{r},\mathbf{r}'  \right) \cdot \text{d}\mathbf{r}'
\label{two-body2}
\end{equation}
so that
\begin{equation}
\Delta E_{(2)}=U\left( \mathbf{r}_\text{A}\right) + U\left( \mathbf{r}_\text{A},\mathbf{r}_\text{B} \right) 
\end{equation}

Note that the single-particle shift $U\left( \mathbf{r}_\text{A}\right) $ contains an infinite contribution which comes from the free Coulomb interaction; however this shift does not depend on the position of the particle and does not lead to any observable force. Note that our results \eqref{one-body} and \eqref{two-body2} remain valid for non-local media where the mode tensor \eqref{mode} involves  a convolution over the non-local permittivity \cite{butcher}. \\

In Eq. \eqref{one-body} the first term represents the classical interaction while the second term is a quantum correction that vanishes in the classical limit $\hbar \to 0$. The single-particle shift represents a correction to the force if the charge is situated in an environment. In general the expressions for such a correction are very complicated, but it is in fact possible to estimate the order of magnitude of this term by direct inspection of Eq.~\eqref{one-body}. Firstly we rewrite \eqref{one-body} in terms of the Compton wavelength $\lambda_\text{A}=2\pi\hbar /(m_\text{A}c)$
\begin{multline}
U\left( \mathbf{r}_\text{A}\right) =  - \frac{q_\text{A}^2}{2 \epsilon_0}\left[  \int\limits_{\mathbf{r}_0}^{\mathbf{r}_\text{A}} \int\limits_{\mathbf{r}_0}^{\mathbf{r}_\text{A}} \text{d}\mathbf{r} \cdot \tens{G}\left( \mathbf{r},\mathbf{r}'  \right) \cdot \text{d}\mathbf{r}' 
-  \frac{  \lambda_\text{A} }{2 \pi^2   } \frac{1}{c}\int\limits_0^\infty  \text{d}\omega \text{Tr}\left\{ \text{Im} ^ \bot \tens{G}\left( \mathbf{r}_A,\mathbf{r}_A,\omega  \right)^ \bot  \right\}    \right]
\end{multline}

We begin by noting that the Green's tensor $\tens{G}\left( \mathbf{r},\mathbf{r}' ,\omega \right)$ typically has an order of magnitude of $r^{-1}$ where $r$ represents the typical distance between the charge and the surface of the medium. Then, Eq.~\eqref{g} tells us that   that the static Green's tensor $\tens{G}\left( \mathbf{r},\mathbf{r}'  \right)$ has an order of magnitude of $r^{-3}$, which in turn means that $\frac{1}{c}\text{Tr} \frac{\text{d}}{\text{d}\omega } \omega^2 \tens{G}\left( \mathbf{r},\mathbf{r}'  \right)$ is of order of magnitude of $r^{-2}$. Hence, the ratio between the second and first term of \eqref{one-body} is of order of magnitude $\lambda_\text{A}/r$. An electron has $\lambda_\text{A} \approx 10^{-12} $m, this should be contrasted with the fact that the description of the surface as a macroscopic body breaks down at distances of around $10^{-9}$m. Thus for all distances for which the basic assumptions of this work hold, the second term of Eq.~\eqref{one-body} can safely be discarded.\\

This analysis is backed up by the detailed evaluation of both terms previously carried out in \cite{barton}, where the problem was analyzed in both the non-retarded and the retarded regimes  for a plasma surface. 
There, it is found that the ratio of the second term to the first term is equal to  $\lambda_\text{A} \lambda_p /(2 \pi \sqrt{2} r^2)$ in the non-retarded regime, and  $\lambda_\text{A}/(\pi r)$ in the retarded regime, where $\lambda_p$ is the plasma wavelength. Since for real materials the plasma wavelength is in the visible range  we recognize also in this case that the first term of \eqref{one-body} dominates the second.\\

Finally, we note that we can  give an interpretation of the classical shift in terms of the work needed to bring the static charges from infinity and assemble them in the required positions. 
The medium-assisted Coulomb electric field produced by the charge A, with position $\textbf{r}_\text{A}$ is:
\begin{equation}
\textbf{E}_\text{A}\left( \mathbf{r}' \right) =   \frac{q_\text{A}}{ \epsilon_0} \int\limits_{\mathbf{r}_0}^{\mathbf{r}_\text{A}} \text{d}\mathbf{r}  \cdot \tens{G}\left( \mathbf{r},\mathbf{r}'  \right) 
\end{equation}
Hence the potential in $\textbf{r}$ due to both charges is:
\begin{equation}
\phi \left( \mathbf{r} \right) =  - \int\limits_{\mathbf{r}_0}^{\mathbf{r}} \text{d} \mathbf{r}' \cdot \left( \mathbf{E}_\text{A}\left( \mathbf{r'} \right) + \mathbf{E}_\text{B}\left( \mathbf{r'} \right) \right)
\end{equation}
The work $W$ required to assemble the charges is 
\begin{equation}
W= \frac{1}{2}\sum\limits_{j = A,B} q_j \phi \left( \mathbf{r}_j \right)\,,
\end{equation}
which coincides with the classical energy shift derived previously in Eq. \ref{energyshift}.

 \section{Energy shift in terms of the scalar Green's function}
\label{GenShiftSection}
The Coulomb interaction depends on the static Green tensor $\tens{G}(\mathbf{r},\mathbf{r}')$ as defined in Eq.~\eqref{g}, which is related to the scalar Green function $g(\mathbf{r},\mathbf{r}')$ from electrostatics defined via \cite{schwinger}:
\begin{equation}
\nabla  \cdot \left[ \epsilon \left( \mathbf{r} \right) \cdot \nabla g\left( \mathbf{r},\mathbf{r}' \right) \right] =  - \delta \left( \mathbf{r} - \mathbf{r}' \right)
\label{greenf}
\end{equation}

 In order to show the connection between the two quantities $\tens{G}(\mathbf{r},\mathbf{r}')$ and $g(\mathbf{r},\mathbf{r}')$ we write the equation of the full, frequency-dependent dyadic Green's tensor $\tens{G}\left( \mathbf{r},\mathbf{r}',\omega  \right)$, see Eq. 	\eqref{FullG}.
Taking the divergence of both sides of $\eqref{FullG}$ and considering the static limit $\omega \to 0$ one has;
\begin{equation}\label{DivHelmholtz}
\nabla  \cdot \left[ \epsilon \left( \mathbf{r} \right) \cdot \tens{G}\left( \mathbf{r},\mathbf{r}' \right) \right] =  - \nabla \delta \left( \mathbf{r} - \mathbf{r}' \right)
\end{equation}
where $\epsilon \left( \mathbf{r} \right)\equiv \epsilon \left( \mathbf{r},\omega\to 0 \right)$ is the static permittivity  and $\tens{G}\left( \mathbf{r},\mathbf{r}' \right)$ the static Green's tensor, see Equation  (\ref{g}).

Applying the operator $\nabla'$ to both sides of Eq.~\eqref{greenf} and comparing the result to Eq.~\eqref{DivHelmholtz} we see that:
\begin{equation}\label{BigGvsSmallG}
\tens{G}\left( \mathbf{r},\mathbf{r}' \right) =  - \nabla \nabla 'g\left( \mathbf{r},\mathbf{r}' \right)
\end{equation}
Using the well-known relation 
\begin{equation}\int\limits_{\mathbf{r}_0}^{\mathbf{r}} \text{d}\mathbf{r}'  \cdot \nabla' f\left( \mathbf{r}' \right) = f\left( \mathbf{r} \right) - f\left( \mathbf{r}_0 \right),\end{equation}
to replace $g$ by $\tens{G}$ in Eqs.~\eqref{one-body} and \eqref{two-body2}, we find:
\begin{align}\nonumber
U\left( \mathbf{r}_\text{A}\right) =&   \frac{q_\text{A}^2}{2 \epsilon_0} g(\mathbf{r}_\text{A},\mathbf{r}_\text{A})  \\ 
U\left( \mathbf{r}_\text{A},\mathbf{r}_\text{B} \right) =&   \frac{q_\text{A} q_\text{B}}{ \epsilon_0} g(\mathbf{r}_\text{A},\mathbf{r}_\text{B}) 
\end{align}
where we have additionally used the property $g\left( \mathbf{r},\mathbf{r}' \right) \to 0$ for $\left| \mathbf{r} - \mathbf{r}' \right| \to \infty $.  In free space we obtain the well-known Coulomb potential since $g^{(0)}(\mathbf{r}_\text{A},\mathbf{r}_\text{B}) = 1/(4\pi \left| \mathbf{r}_\text{A} - \mathbf{r}_\text{B} \right|)$.
In general the Green's function is the sum of the free-space contribution $g^{(0)}$ and the scattering part $g^{(1)}$ which accounts for reflections from any surfaces that may be present. In the rest of this work we subtract from $U\left( \mathbf{r}_\text{A}\right)$ the divergent free-space contribution as our main focus is the corrections to thus stemming from the electromagnetic environment, thus we will work with the following pair of equations;
\begin{align}
\label{egreenA} U^{(1)}\left( \mathbf{r}_\text{A}\right) =&   \frac{q_\text{A}^2}{2 \epsilon_0} g^{(1)}(\mathbf{r}_\text{A},\mathbf{r}_\text{A}), \\ 
U\left( \mathbf{r}_\text{A},\mathbf{r}_\text{B} \right) =&   \frac{q_\text{A} q_\text{B}}{ \epsilon_0} g(\mathbf{r}_\text{A},\mathbf{r}_\text{B}). 
\label{egreenAB}
\end{align}
The Green's function  $g(\mathbf{r},\mathbf{r}')$ represents the propagator of the field, describing the amplitude for a photon emitted at $\mathbf{r}'$ to be absorbed at $\mathbf{r}$. Hence Eq.~\eqref{egreenA} represents an emission and absorption of a photon by the charge, with a reflection from the surface.  In fact the reflected photon can be thought of as emitted by a fictitious image charge.
Eq.~\eqref{egreenAB} represents a  back-and-forth excitation exchange between the two charges (with possible reflection).\\

For transitionally invariant media the Green's function depends only on the difference between the two points: $g(\mathbf{r}_\text{A},\mathbf{r}_\text{B})=g(\mathbf{r}_\text{A}-\mathbf{r}_\text{B})$. In this kind of system the forces acting on the two charges would be equal and opposite, which is just a consequence of the action-reaction principle. However when the translational invariance is broken and local-field corrections are taken into account the two forces are not equal and opposite. This is not a violation of action-reaction since the interface between two different media takes some momentum to restore the balance.

\section{Local-field  corrections to the Coulomb interaction}\label{LocalFieldSection}
\begin{figure}[ht]
   \centering
   \includegraphics[scale=0.85]{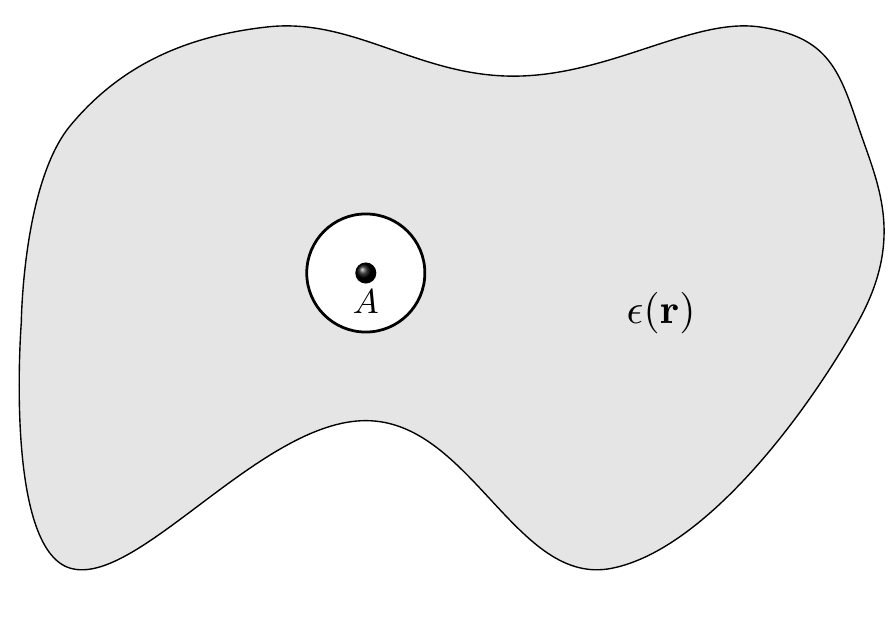}
   \caption{Cavity model: the guest charges in a medium are separated by small, vacuum-filled spherical cavities of radius $R_c$. }
    \label{fig3}
   \end{figure}

In section \eqref{GenShiftSection} we derived the Coulomb interaction between two charges placed within a generic environment. It was assumed that the local electromagnetic field acting on the two particles is the macroscopic field. However this assumption is not satisfied in optically dense media, where local field corrections are important \cite{sambale}. One way to introduce the local-field correction is via the real-cavity model. There, the charges are considered to be surrounded by small, empty, spherical cavities and hence well-separated from the neighbouring atoms of the media as shown in Fig.~(\ref{fig3}).
 If we suppose that the charge A is in such a cavity and that the surrounding medium is infinite, the electric field according to Gauss's law reads:
\begin{equation}
\mathbf{E}\left( \mathbf{r} \right) =  - \frac{q_A}{4\pi \epsilon _0\epsilon }\nabla \frac{1}{\left| \mathbf{r} - \mathbf{r}_\text{A} \right|}
\end{equation}
where $r>R_c$ ($R_c$ radius of the cavity).
 \\
   This field can be expressed in terms of the Green's function of an infinite body without the cavity $g^{\left( 0 \right)}$:
   \begin{equation}
\mathbf{E}\left( \mathbf{r} \right) =  - \frac{q_A}{\epsilon _0}\nabla g^{\left( 0 \right)}\left( \mathbf{r},\mathbf{r}_\text{A} \right)
\end{equation}
However the medium can have a finite size and the electric field can be reflected from the outer surface of the medium. The reflection is mathematically described by the scattering Green's function $g^{\left( 1 \right)}$, hence the reflected field reads:
  \begin{equation}
\mathbf{E}^{\text{ref}}\left( \mathbf{r} \right) =  - \frac{q_A}{\epsilon _0}\nabla g^{\left( 1 \right)}\left( \mathbf{r},\mathbf{r}_\text{A} \right)
\end{equation}
The electric field will be transmitted to the interior to the cavity and eventually modified by polarization effects. According to \cite{greiner} the interior electric field, for small cavity radius, reads:
\begin{equation}
\mathbf{E}^{\text{ref}}_<\left( \mathbf{r} \right) =  -\frac{3\epsilon \left( \mathbf{r}_\text{A} \right)}{2\epsilon \left( \mathbf{r}_\text{A} \right) + 1} \frac{q_A}{\epsilon _0}\nabla g^{\left( 1 \right)}\left( \mathbf{r},\mathbf{r}_\text{A} \right)
\end{equation}
The local-field corrected force acting on the charge A reads:
\begin{equation}
\mathbf{F}_A=q_A\mathbf{E}^{\text{ref}}_<\left( \mathbf{r}_\text{A} \right) =  -\frac{3\epsilon \left( \mathbf{r}_\text{A} \right)}{2\epsilon \left( \mathbf{r}_\text{A} \right) + 1} \frac{q_A^2}{\epsilon _0}\nabla g^{\left( 1 \right)}\left( \mathbf{r},\mathbf{r}_\text{A} \right)|_{\mathbf{r}=\mathbf{r}_\text{A}}
\label{egreenAcorr}
\end{equation}
 We now focus on the local-field corrected interaction between two charges. Charge B will produce some field which is experienced by charge A.
 The electric field produced by B reads:
 \begin{equation}
\mathbf{E}^{B}\left( \mathbf{r} \right) =  - \frac{q_B}{\epsilon _0}\nabla g \left( \mathbf{r},\mathbf{r}_B \right)
\end{equation}
which has two contributions: these are the field produced directly by charge B, and the field reflected from the outer surface respectively. Hence the local-field corrected electric field inside the cavity of the charge A reads:
   \begin{equation}
\mathbf{E}^{B}_<\left( \mathbf{r} \right) =  -\frac{3\epsilon \left( \mathbf{r}_\text{A} \right)}{2\epsilon \left( \mathbf{r}_\text{A} \right) + 1} \frac{q_B}{\epsilon _0}\nabla g \left( \mathbf{r},\mathbf{r}_B \right)
\end{equation}
The force experienced by the charge A is then:
\begin{equation}
\mathbf{F}_A=q_A\mathbf{E}^{B}_<\left( \mathbf{r}_\text{A} \right) =  -\frac{3\epsilon \left( \mathbf{r}_\text{A} \right)}{2\epsilon \left( \mathbf{r}_\text{A} \right) + 1} \frac{q_A q_B}{\epsilon _0}\nabla_A g\left( \mathbf{r}_\text{A},\mathbf{r}_B \right)
\label{egreenABcorr}
\end{equation}
Local-field corrections lead to enhancement of the force. For example, water has a static permittivity of approximately 80 leading to an enhancement factor of about $1.49$.

\section{Applications}
To demonstrate the use of our general results \eqref{egreenA}-\eqref{egreenAB} and their generalizations \eqref{egreenAcorr}-\eqref{egreenABcorr} including local-field effects, we apply them to several geometries.

\subsection{Homogeneous non-local medium}
As an example we first consider a translationally invariant medium, which is also  spatially dispersive. Spatial dispersion is the dependence of the permittivity on the wave vector (rather than simply its magnitude), which means that the electric induction $\textbf{D}$ at some point is caused by the electric field $\textbf{E}$ at one or more displaced points. In a bulk medium the Green's function must be translation-invariant, which means that it can only depend on the difference between the coordinates:
\begin{equation}
\tens{G}\left( \mathbf{r},\mathbf{r}',\omega  \right) = \tens{G}\left( \mathbf{r} - \mathbf{r}',\omega  \right)
 \end{equation} 
In this case the Green's dyadic is often Fourier transformed:
\begin{equation}
\tens{G}\left( \mathbf{r} - \mathbf{r}',\omega  \right) = \left( 2\pi  \right)^{-3}\int \text{d}^3 k \text{e}^{\text{i}\mathbf{k} \cdot \left( \mathbf{r} - \mathbf{r}' \right)} \tens{G}\left( \mathbf{k},\omega  \right)
 \label{green0}
\end{equation}
where the Fourier transform can be given in terms of the transverse and longitudinal permittivities, $\epsilon _ \bot \left( k,\omega  \right)$ and $\epsilon _\parallel \left( k,\omega  \right)$ \cite{horsley}:
\begin{equation}
\tens{G}\left( \mathbf{k},\omega  \right) =  - \left[ \frac{\omega ^2}{c^2} \epsilon _ \bot \left( k,\omega  \right) - k^2 \right]^{ - 1}\left( \mathbb{I} - \frac{\mathbf{k}  \mathbf{k}}{k^2} \right)
 - \frac{c^2}{\omega ^2k^2\epsilon _\parallel \left( k,\omega  \right)} \mathbf{k} \mathbf{k}
 \label{green1}
\end{equation}
Here $\mathbb{I}$ is the identity matrix.
Hence the Green's function reads:
\begin{equation} \label{Grr}
g\left( \mathbf{r},\mathbf{r}' \right) =   \frac{1}{\left( 2\pi  \right)^3}\int \text{d}^3 k\frac{1}{k^2\epsilon _\parallel \left( k,0 \right)}\text{e}^{\text{i}\mathbf{k} \cdot \left( \mathbf{r} - \mathbf{r}'\right)}
\end{equation}
Substituting the Fourier-transformed Green function \eqref{two-body2} into the two-body Coulomb potential given by Eq.~\eqref{egreenAB}, we find:
\begin{equation}\label{NonLocalEnergyShift}
U\left( \mathbf{r}_\text{A},\mathbf{r}_\text{B} \right) = \frac{q_\text{A}q_\text{B}}{\left( 2\pi  \right)^3\epsilon _0}\int \text{d}^3 k\frac{1}{k^2\epsilon _\parallel \left( k,0  \right)}\text{e}^{\text{i}\mathbf{k} \cdot \left( \mathbf{r}_\text{A} - \mathbf{r}_\text{B} \right)} =
 \frac{q_\text{A}q_\text{B}}{4\pi ^2\epsilon _0 \text{i}}\int\limits_{ - \infty }^\infty  \text{d}k \frac{1}{\epsilon _\parallel \left( k,0 \right)}\frac{\text{e}^{\text{i}kr}}{kr}
\end{equation}
where $\textbf{r}\equiv\textbf{r}_\text{A}-\textbf{r}_\text{B}$.

 The longitudinal and transverse nonlocal permittivity of a real medium can be described by the hydrodynamic Drude model, which can be considered a limiting case of the Hopfield model \cite{raza,hopfield,gale}:
 \begin{equation}\label{LongPerm}
\epsilon _\parallel \left( k,\omega  \right) = 1+ \frac{\tilde \omega _p^2}{\omega _0^2 -\omega \left( \omega  + \text{i}\Gamma  \right)}+ \frac{\omega_p^2}{ \beta ^2 k^2-\omega \left( \omega  + \text{i}\gamma  \right)} .
 \end{equation}
The first term represents the dielectric function for bound electrons and the second one that for conduction electrons. The plasma frequencies are defined by $\omega _p = \sqrt {ne^2/m_e\epsilon _0} $, $\tilde \omega _p = \sqrt {Ne^2/m\epsilon _0} $, where $n,m_e$ are respectively the number density and effective mass of the free electrons and the $N$ is the density of bound electrons. Finally, $\omega_0$ is a transition frequency and $\Gamma, \gamma>0$ are damping constants that govern absorption in the medium. For a free electron gas $\beta^2=3v_F^2/5$, with $v_F$ the Fermi velocity \cite{kittel}. 

Inserting the longitudinal permittivity \eqref{LongPerm} into the energy shift \eqref{NonLocalEnergyShift} and evaluating the residua of the poles in the upper part of the complex plane we find; 
\begin{equation}
U\left( r \right) = \frac{q_\text{A}q_\text{B}}{4\pi \epsilon _0\epsilon}\frac{\text{e}^{ - k_sr}}{r}
\end{equation}
with $\epsilon=1+ {\tilde \omega _p^2}/{\omega _0^2 }$ and $k_s = {\omega _p}/(\beta \sqrt {\epsilon} )$. This is a screened Coulomb potential where both bound and free electrons contribute to the screening \cite{ashcroft,kittel}.

 \subsection{Dielectric slab}
 In the previous section we considered a homogeneous medium. However in many interesting cases translational symmetry is broken, for example in planar multilayer systems. \\
 
\begin{figure}[ht]
   \centering
   \includegraphics[scale=0.85]{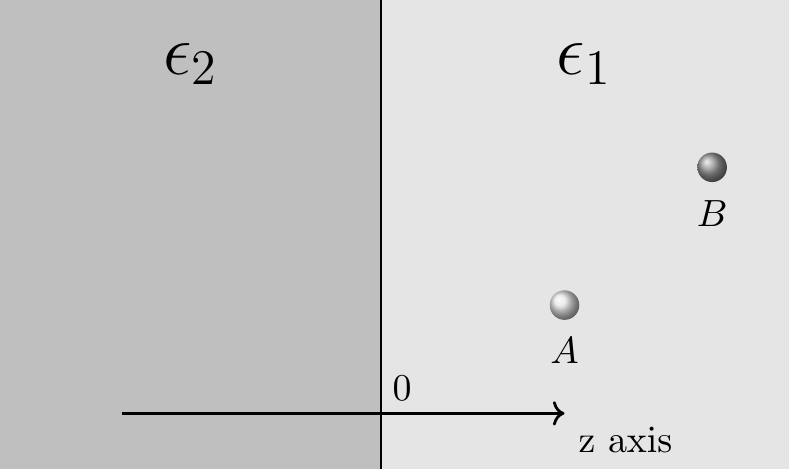}
   \caption{Two point charges near a semi infinite dielectric.}
    \label{fig4}
   \end{figure}

 Here we consider two point charges  embedded in a semi-infinite dielectric medium of dielectric constant  $\epsilon_1$, which has a planar interface with a material of dielectric constant  $\epsilon_2$ (see Fig.~\ref{fig4})
   
The Green's function satisfies Eq.~\eqref{greenf},
with the boundary conditions that $g$ and the normal component of the displacement vector $\mathbf{D}$ are continuous across the interface between two media.
Its expression for source position $z'>0$ reads \cite{schwinger}:
\begin{equation}  g(\mathbf{r},\mathbf{r}')=\begin{cases}
  \frac{1}{\epsilon _1}\frac{1}{4\pi\left| \mathbf{r} - \mathbf{r}' \right|} + \frac{1}{\epsilon _1}\frac{\epsilon _1 - \epsilon _2}{\epsilon _1 + \epsilon _2}\frac{1}{4\pi\left| \mathbf{r} - \mathbf{r}{'}^\star \right|},& \text{if } z>0\\
            \frac{2}{\epsilon _1 + \epsilon _2}\frac{1}{4\pi\left| \mathbf{r} - \mathbf{r}' \right|},         & \text{if } z<0
\end{cases}\end{equation}
where $\mathbf{r}^\star=(x,y,-z)$ is the position of an image charge that corresponds to a real charge placed at $\mathbf{r}$. Inserting this into \eqref{egreenAB} we obtain  the following local-field corrected Coulomb interaction between the two charges:
\begin{equation}
U\left( \mathbf{r}_\text{A},\mathbf{r}_\text{B} \right) = \frac{q_\text{A}q_\text{B}}{4\pi \epsilon _0\epsilon _1} 
 \bigg( \frac{1}{\left| \mathbf{r}_\text{A} - \mathbf{r}_\text{B} \right|} 
-\frac{\epsilon _2 - \epsilon _1}{\epsilon _2 + \epsilon _1}\frac{1}{\left| \mathbf{r}_\text{A} - \mathbf{r}_\text{B}^ \star  \right|}\bigg)
\end{equation}
where $\textbf{r}_\text{B}^\star =(x_\text{B},y_\text{B},-z_\text{B})$ is the position of the image of charge B. The first term represents the free Coulomb interaction, while the second one represents the interactions between the real charge $A$ and the image charge $B^\star$. 
The local-field corrected force on A reads:
\begin{equation}
\mathbf{F}_A=- \frac{q_\text{A}q_\text{B}}{4\pi \epsilon _0\epsilon _1} \frac{3 \epsilon_1}{2\epsilon_1+1} 
\nabla_A \bigg( \frac{1}{\left| \mathbf{r}_\text{A} - \mathbf{r}_\text{B} \right|} 
-\frac{\epsilon _2 - \epsilon _1}{\epsilon _2 + \epsilon _1}\frac{1}{\left| \mathbf{r}_\text{A} - \mathbf{r}_\text{B}^ \star  \right|}\bigg)
\end{equation}

We can describe as well the single-particle Coulomb term:
\begin{equation}
U^{(1)}(\mathbf{r}_\text{A} ) =  - \frac{q_\text{A}^2}{8\pi \epsilon _0\epsilon _1}    \frac{\epsilon _2 - \epsilon _1}{\epsilon _2 + \epsilon _1}\frac{1}{\left| \mathbf{r}_\text{A} - \mathbf{r}_\text{A}^\star \right|}=- \frac{q_\text{A}^2}{16\pi \epsilon _0\epsilon _1}    \frac{\epsilon _2 - \epsilon _1}{\epsilon _2 + \epsilon _1}\frac{1}{z_A}
\end{equation}
and the related corrected force:
\begin{equation}
\mathbf{F}_A=- \frac{q_\text{A}^2}{16\pi \epsilon _0\epsilon _1}    \frac{\epsilon _2 - \epsilon _1}{\epsilon _2 + \epsilon _1} \frac{3 \epsilon_1}{2\epsilon_1+1}  \frac{\hat z}{z_A^2}.
\end{equation}
This can be interpreted in terms of the interaction between the real charge A and its image. The factor $1/2$ arises because the the image charge is not a real charge, but rather an induced one.

We also briefly comment on the the case where one charge is embedded in a medium of dielectric constant $\epsilon_1$ and the other charge embedded in the medium of dielectric constant $\epsilon_2$. In this case:
\begin{equation}
U\left( \mathbf{r}_\text{A},\mathbf{r}_\text{B} \right) = 
 \frac{q_\text{A}q_\text{B}}{2\pi \epsilon _0} \frac{1}{\epsilon _1 + \epsilon _2}    \frac{1}{\left| \mathbf{r}_\text{A} - \mathbf{r}_\text{B} \right|}
\end{equation}
In particular if one medium is a perfect conductor (e.g. $\epsilon_1 \to \infty$), the two charges do not interact, since any photon emitted by one charge is completely reflected by the interface and does not reach the other charge.

\subsection{Cavity.}
\begin{figure}[ht]
   \centering
   \includegraphics[scale=0.85]{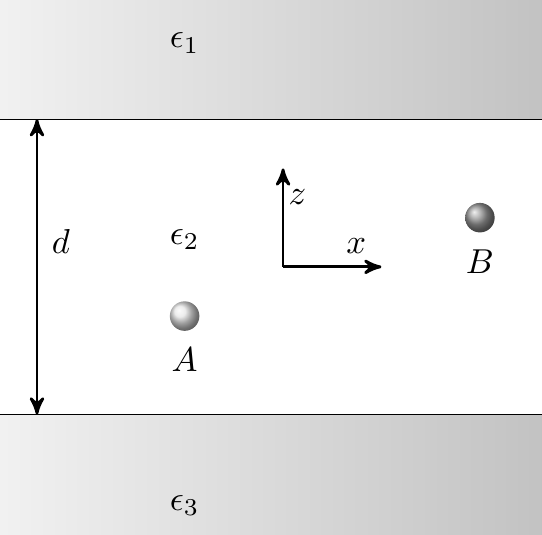}
   \caption{Two charges inside a dielectric cavity.}
    \label{fig5}
   \end{figure}

Our next example is the Coulomb interaction between two charges in a cavity. We suppose that two charges are embedded in a medium with dielectric constant $\epsilon_2$ and is near two other parallel semi-infinite dielectrics with relative permittivity $\epsilon_1$ and $\epsilon_3$. The plane $z=0$ is equidistant from the two dielectric surfaces, which are separated by a distance $d$, as shown in fig.~\ref{fig5}

\begin{figure}[h!]
   \centering
   \includegraphics[scale=0.65]{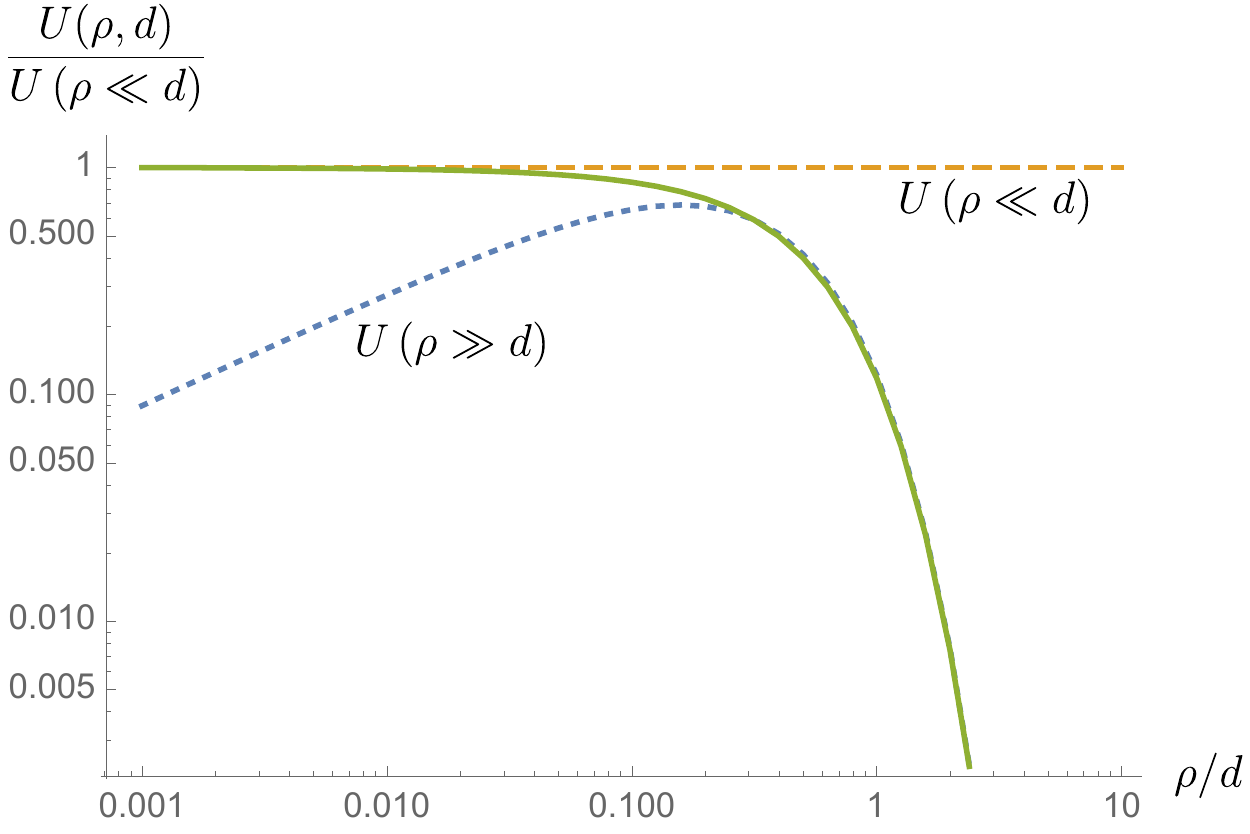}
   \caption{Coulomb interaction between two charges in vacuum at the midpoint of a perfectly reflecting cavity. At small distances the interaction is equal to the standard Coulomb result \eqref{Smallrho}, and at large distances it is equal to the asymptotic result \eqref{LargeDistanceAsymptotic} where the interaction is exponentially supressed.}
    \label{ExponentialTransitionFig}
   \end{figure}

At the midpoint between the plates ($z=z'=0$), the Green's function (see \ref{Appendix}) reads
\begin{equation}\label{g2Slabs}
g_2 =\frac{1}{4\pi \epsilon _2}\int\limits_0^\infty  \text{d}k  \frac{\left( 1 - \text{e}^{ - k d }R_1 \right)\left( 1 - \text{e}^{ - kd}R_3 \right)}{\left( 1 - \text{e}^{ - 2kd}R_1R_3 \right)}J_0\left( k\rho  \right)
\end{equation}
where $J_0$ is the zeroth-order Bessel function of the first kind, and $R_1$ and $R_3$ are the reflection coefficients for the two media bounding region 2;
\begin{align}
R_1 = \frac{\epsilon _1 - \epsilon _2}{\epsilon _1 + \epsilon _2}, \quad R_3 = \frac{\epsilon _3 - \epsilon _2}{\epsilon _3 + \epsilon _2}
\label{R1r3}
\end{align}
We can expand the denominator of \eqref{g2Slabs} in a power series $\left( 1 - \text{e}^{ - 2kd}R_1R_3 \right)^{ - 1} = \sum\limits_{n = 0}^\infty  \left( R_1R_3\text{e}^{ - 2kd} \right)^n $, which can then be integrated term-by-term
\begin{equation}\label{CavityGSum}
g_2 = \frac{1}{4\pi \epsilon _2\rho } + \frac{1}{4\pi \epsilon _2}\sum\limits_{n = 1}^\infty  \frac{2\left( R_1R_3 \right)^n}{\sqrt{d^2\left( 2n \right)^2 + \rho ^2}}  - \frac{1}{4\pi \epsilon _2}\sum\limits_{n = 0}^\infty  \frac{\left( R_1R_3 \right)^n\left( R_1 + R_3 \right)}{\sqrt{ d^2\left( 2n + 1 \right)^2 + \rho ^2 }} 
\end{equation}
Each term here can be interpreted as what one would have got using the image method, but the method outlined here and in \ref{Appendix} is more convenient as in this case we would have to deal with an infinite series of images corresponding to multiple reflections. 
An asymptotic limit can be derived when the left and right surfaces are perfect conductors: $R_1=R_3=1$, and the charges are separated by a much larger distance than that between the plates; $\rho \gg d$  \cite{Pumplin,Melo}:
\begin{equation}
g_2 = \frac{1}{4\pi \epsilon _2}\sqrt {\frac{8}{\rho d}} \text{e}^{ - \pi \rho /d}
\end{equation}
Since both charges are equidistant from the conducting surfaces ($z=z'=0$) the interaction between the charges and the surfaces vanishes, as one would expect from symmetry considerations. However the interaction between the two charges does not vanish. In particular for $ \rho \gg d$, according to Eq.~(\ref{egreenAB}) the interaction reads:
\begin{equation}\label{LargeDistanceAsymptotic}
U\left( \rho \gg d \right) = \frac{q_\text{A}q_\text{B}}{4\pi \epsilon _0 \epsilon _2}   \sqrt {\frac{8}{\rho d}} \exp \left(  - \frac{\pi \rho}{d} \right)
\end{equation}
Hence the cavity exponentially suppresses the Coulomb interaction at large distances. At small $\rho$ (or equivalently large $d$) the interaction is just the Coulomb potential between two isolated charges
\begin{equation}\label{Smallrho}
U\left( \rho \ll d \right) =\frac{q_\text{A}q_\text{B}}{4\pi\epsilon_0 \rho},
\end{equation}
which we plot in fig.~\ref{ExponentialTransitionFig},
alongside the large-distance asymptotic result \eqref{LargeDistanceAsymptotic} and the result obtained from numerical evaluation of  \eqref{egreenAB} using \eqref{CavityGSum}. Finally, if the region inside the cavity has $\epsilon_2\neq 1$, the local-field corrected force reads:
\begin{equation}
\mathbf{F}_A =\frac{q_\text{A}q_\text{B}}{4\pi \epsilon _0 \epsilon _2} \frac{3\epsilon_2}{2\epsilon_2+1}  \frac{\sqrt 2 }{\left( \rho d \right)^{3/2}}\exp \left(  - \frac{\pi \rho }{d} \right)\left( d + 2\pi \rho  \right)
\end{equation}

\subsection{Plate with a hole}
\begin{figure}[ht]
   \centering
   \includegraphics[scale=0.85]{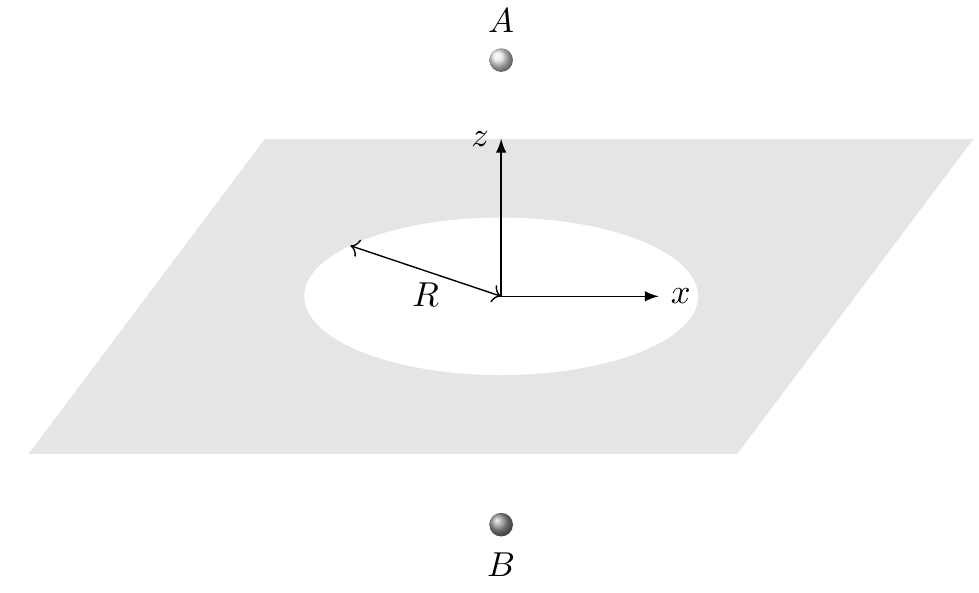}
   \caption{Two charges are located near a perfect conducting plate with a hole of radius $R$. The symmetry axis of the plate is the $\hat z$ axis.}
    \label{fig6}
   \end{figure}

We now consider a more involved example, namely a perfect conducting plate with a hole of radius $R$, as shown in \ref{fig6}.
Analogous problems involving the plate with a hole have been considered in the literature \cite{jackson,smythe}.\\

\begin{figure}[ht]
   \centering
   \includegraphics[scale=0.6]{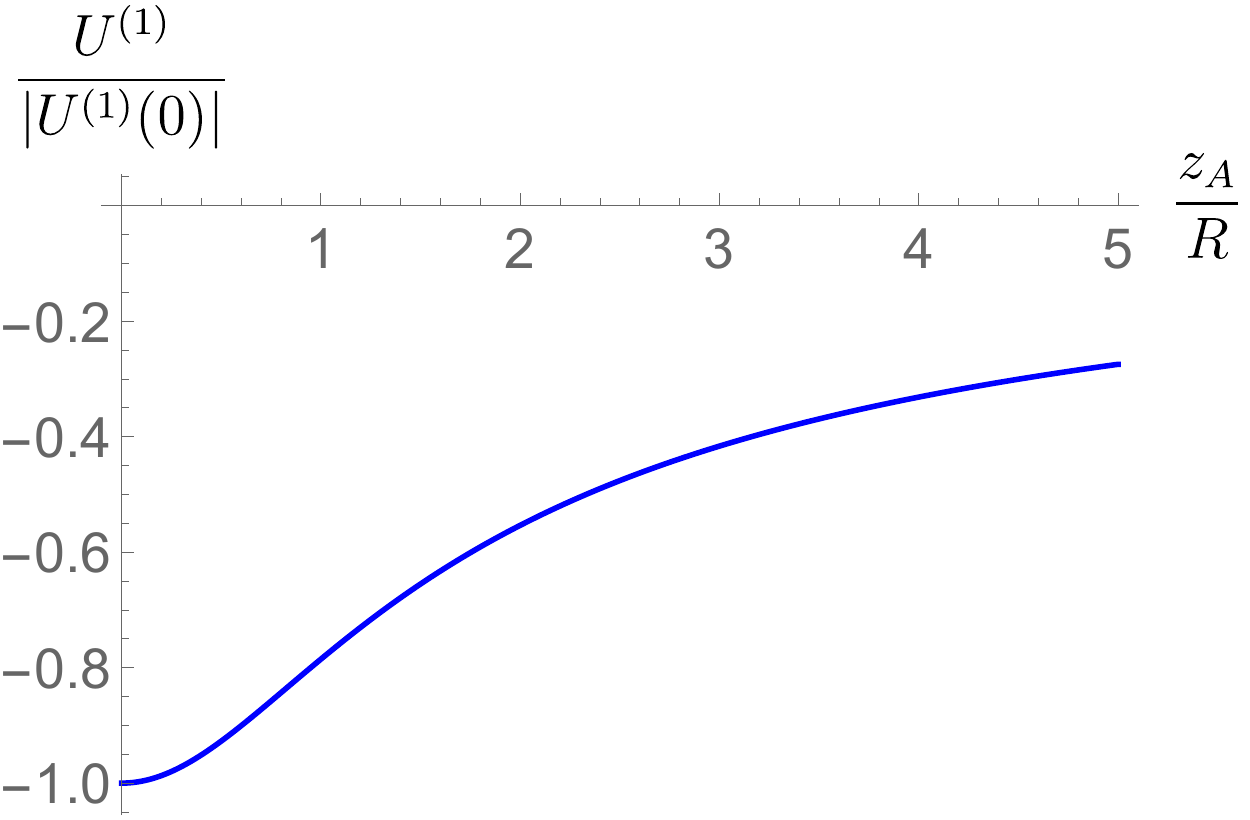}
   \caption{Coulomb interaction between an on-axis electron and the plate. The interaction is scaled with respect to the function $U_{free}^{\left( 1 \right)} = \mathop {\lim }\limits_{R \to \infty } U^{\left( 1 \right)} =- q_\text{A}^2/8\pi ^2\epsilon _0R$. }
    \label{fig7}
   \end{figure}

 This case is interesting from a technical point of view since it is not obvious how to locate the image charges to satisfy the boundary conditions, and interesting from an applied point of view due to its relevance to membranes. A hole or pore in the membrane of a biological cell can allow a variety of possibly-ionised particles to enter \cite{pont}. The required Green's function may be obtained the Kelvin inversion from the Green's function of the semi-infinite half-plane \cite{eberlein,eberlein2}. However the expression obtained in \cite{eberlein} is valid for $\mathbf{r}$ and $\mathbf{r}'$ lying on the same side of the plate: $z,z'>0$. Here we generalize the Green's function to include the case $z>0,z'<0$ as well, finding:
\begin{equation}\label{GFPlateWithHole}
   g\left( \mathbf{r},\mathbf{r}' \right) = \frac{1}{8\pi }\Bigg\{ \frac{1}{D_ - }\left[ 1 + \frac{2\lambda _ - }{\pi }\arctan \left( \frac{F_ - }{D_ - } \right) \right] 
   -\frac{1}{D_ + }\left[ 1 + \frac{2\lambda  _ + }{\pi }\arctan \left( \frac{F_ + }{D_ + } \right) \right] \Bigg\}
   \end{equation}
   where:
    \begin{align}\nonumber
   F_ \pm  =& \frac{1 }{\sqrt 2R}\Big\{ \left( \rho ^2 + z^2 - R^2 \right)\left( \rho {'^2} + z{'^2} - R^2 \right)\\ \nonumber
    &\mp 4R^2 + \sqrt {\left( z^2 + \left( \rho  - R \right)^2 \right)\left( z^2 + \left( \rho  + R \right)^2 \right)}\\ 
& \times \sqrt {\left( z{'^2} + \left( \rho ' - R \right)^2 \right)\left( z{'^2} + \left( \rho ' + R \right)^2 \right) } \Big\}^{1/2}\\
D_ \pm  =& \sqrt {\rho ^2 + \rho {'^2} - 2\rho \rho '\cos \left( \phi  - \phi ' \right) + \left( z \pm z' \right)} \\\nonumber
     \lambda _ +=&\begin{cases}
\operatorname{sgn} \Big[ z'\left( \rho ^2 + z^2 - R^2 \right)\\
 + z\left( \rho {'}^2 + z{'}^2 - R^2 \right) \Big],& \text{if } z>0,z'>0\\
    -1,& \text{if } z>0,z'<0
\end{cases}\\
\lambda  _ -=&\begin{cases}
1,& \text{if } z>0,z'>0\\
    \operatorname{sgn} \Big[ z'\left( \rho ^2 + z^2 - R^2 \right)\\
 - z\left( \rho {'}^2 + z{'}^2 - R^2 \right) \Big],& \text{if } z>0,z'<0
\end{cases}\end{align}
and $(\rho,\phi,z)$ represents the coordinates of $\mathbf{r}$ in a cylindrical system where the symmetry axis of the hole is the $z$ axis and the origin is at the centre of the hole.
We will initially focus on the case where both charges are on the symmetry axis, and firstly quote the single-particle result found via Eq.~\eqref{egreenA}
     \begin{equation}
   U^{(1)}\left( \mathbf{r}_\text{A} \right) =  - \frac{q_\text{A}^2}{32\pi \epsilon _0z_\text{A}} + \frac{q_\text{A}^2}{16\pi ^2\epsilon _0z_\text{A}}\arctan \left[ \frac{R}{2z_\text{A}} - \frac{z_\text{A}}{2R} \right]
   \end{equation}
   The resulting interaction is shown in Fig. \ref{fig7}, where we have scaled with respect to the interaction for $z_A=0$ ($U^{\left( 1 \right)} (0)=  - \frac{q_\text{A}^2}{8\pi ^2\epsilon _0R}$.)
  The interaction is always attractive and vanishes in the limit $z \to 0$ for finite $R$,which is also clear from symmetry considerations. In the full-plate case ($R=0$) the force diverges as $z\to 0$ since there we can construct the image charge which approaches the real charge. Hence in the limit $z_\text{A} \ll R $, for $d$ finite, the interaction with the surface can be neglected and we can focus only on the medium-assisted interaction between the two charges. \\
   
\begin{figure}[ht]
   \centering
   \includegraphics[scale=0.5]{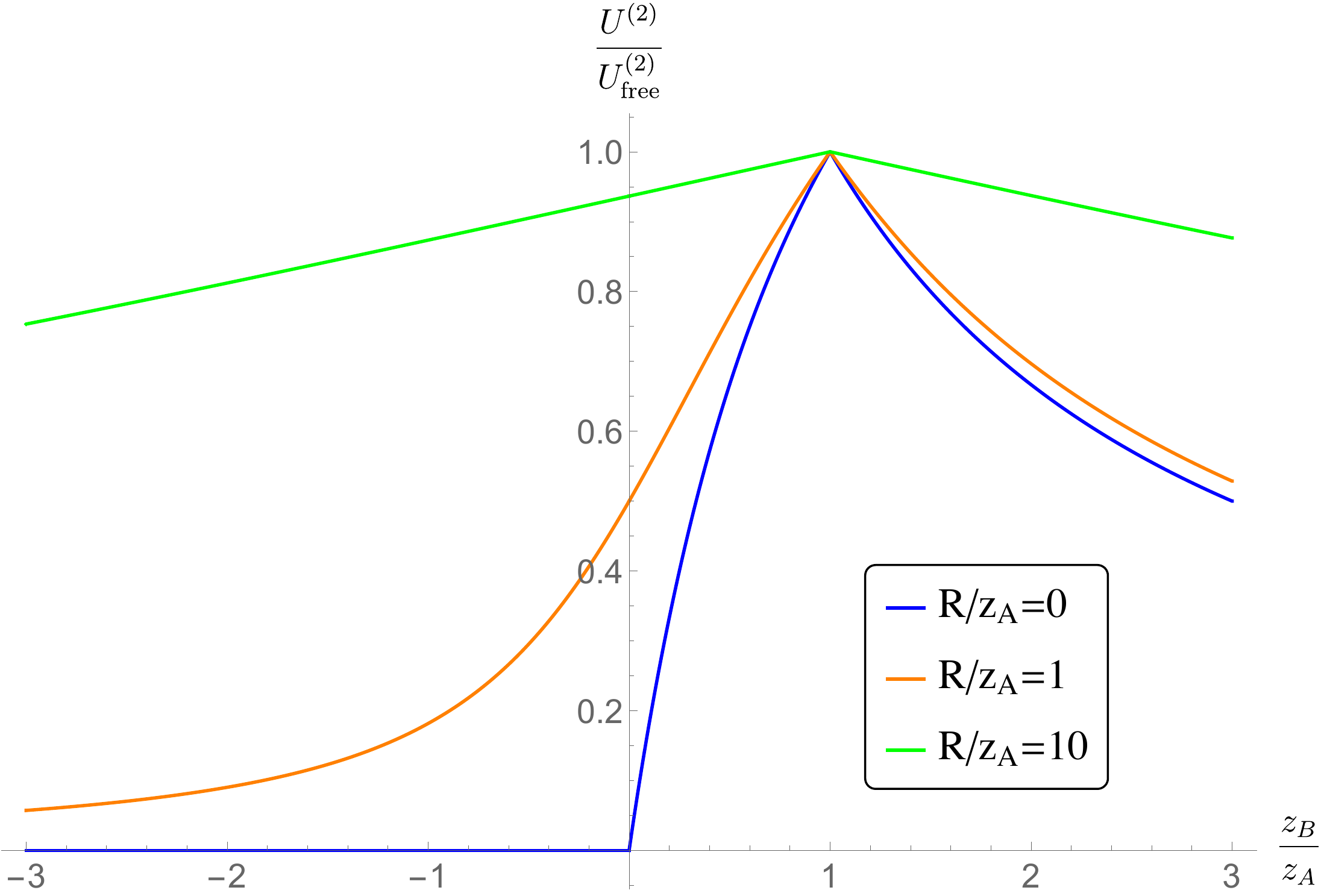}
   \caption{Coulomb interaction between two on-axis charges assisted by the plate. The interaction is scaled with respect to the free interaction $U_{\text{free}}^{\left( 2 \right)} = 1/(4\pi \left| z_\text{A} - z_\text{B} \right|)$. We consider three different values of the radius: $R/z_\text{A}=0,1,10$.}
    \label{fig8}
   \end{figure}

 To calculate the interaction potential between the charges we suppose that one charge in one side of the plate and we vary the position of the other charge.
 If  the two charges are lying on the $z$ axis on opposite sides of the plate: $z>0,z'<0$ and $R \to 0$ we have $F_\pm /D_\pm \to +\infty$ and $\lambda_\pm=-1$ and meaning that the interaction vanishes. This is to be expected since the plate is a perfect conductor, so any photon emitted by one charge is completely reflected by the surface, so the charges do not see each other. If $R$ is finite a photon can travel from one side to to the other, so we expect a non-vanishing interaction. In Fig.~\ref{fig8} we plot the this interaction energy for different value of the radius of the hole. The interaction is scaled with respect to the free interaction $U_{\text{free}}^{\left( 2 \right)} = \frac{1}{4\pi \left| z_\text{A} - z_\text{B} \right|}$ .
For a finite hole radius $R$ there is a non-vanishing weak interaction also when $z_\text{B}<0$.\\

\begin{figure}[ht]
   \centering
   \includegraphics[scale=1]{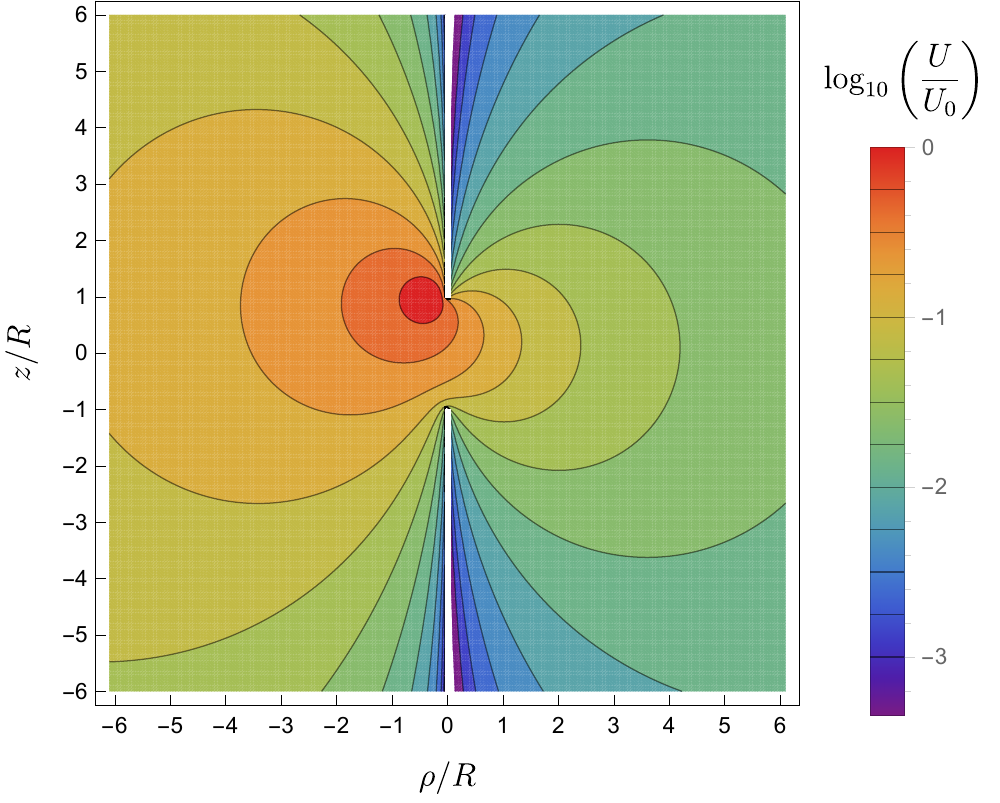}
   \caption{Cross-section of the Coulomb interaction of two charges positions near a plate with a hole. All distances are measured in units of the hole radius $R$, and the position of one charge is taken as $z=-0.15 R$, $\rho = R$. The colours represent the potential felt by the other charge at that point. The interaction is scaled to be in units of the free space result $U_0 = {1}/(4\pi |\mathbf{r}-\mathbf{r}_0|)$, and the two charges are assumed to be lying in the plane shown in the figure (i.e. $\phi'=\phi$)}
    \label{fig9}
   \end{figure}

To demonstrate the power and general applicability of our method we consider now the interaction between charges $A$ and $B$ when each charge may be located at any position. Directly using \eqref{egreenAB} in \eqref{GFPlateWithHole} we can produce the full three-dimensional interaction potential, which we show a slice of in fig.~\eqref{fig9};

\subsection{Interaction between a charged particle and a polarizable molecule}
\begin{figure}[ht]
   \centering
   \includegraphics[scale=0.65]{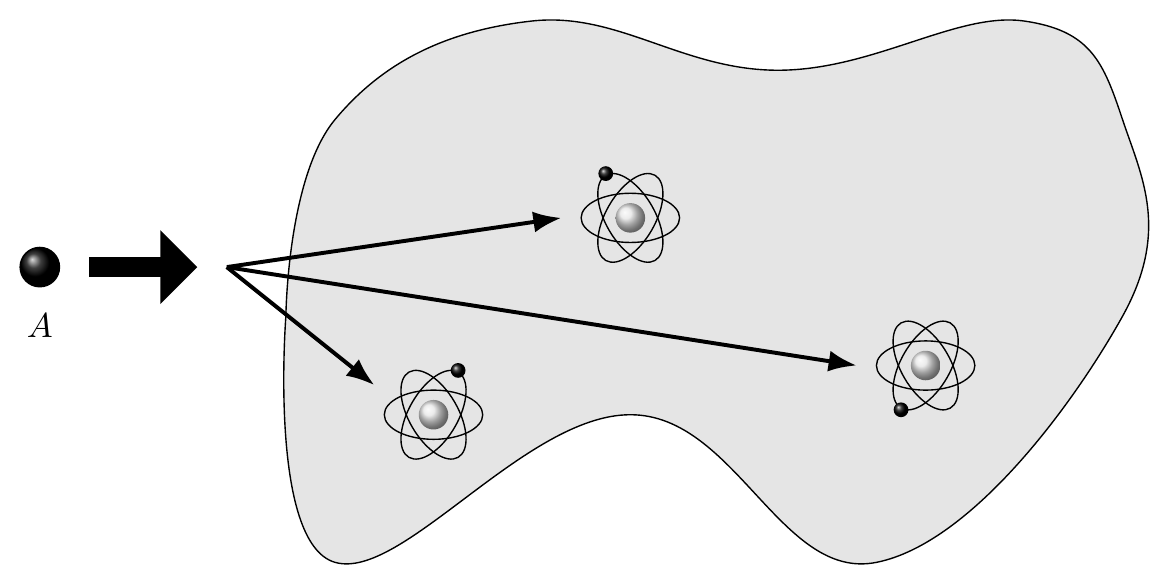}
   \caption{Microscopic interpretation of the Coulomb interaction between a molecule and a polarizable body. }
    \label{fig10}
   \end{figure}

Finally we consider what happens when a charged particle is placed near a polarizable medium, in which case the Coulomb field of the charged particle will polarize the molecules that constitute the medium. In the dilute limit the Coulomb interaction between the charge and the surface can be described in microscopic terms, as arising from the Coulomb interaction of the charge and the individual molecules, as shown in Fig. \ref{fig10}.
It is well-known that for a dilute media a Born expansion of the scattering Green tensor can be performed:
\begin{equation}
\tens{G}^{(1)}\left( \mathbf{r},\mathbf{r}',\omega  \right)=
\mu _0\omega ^2\int \text{d}^3r_\text{B}\,\eta(\mathbf{r}_\text{B}) \tens{G}^{(0)}\left(
\mathbf{r},\mathbf{r}_\text{B},\omega  \right)  \cdot \bm{\alpha}_\text{B}\left(
\omega \right) \cdot \tens{G}^{(0)}\left( \mathbf{r}_\text{B},\mathbf{r}',\omega
\right) + ...
\end{equation}
 where $\eta$ is the number density,  $\tens{G}^0$ is a bulk Green's tensor of a medium considered as a `background' and the integration is over the volume of the polarizable medium. Using Eq.~\eqref{g} we can write the Born expansion in terms of the static Green's function;
\begin{equation}
\tens{G}^{(1)}\left( \mathbf{r},\mathbf{r}'  \right)=
\frac{1}{\epsilon_0}\int \text{d}^3r_\text{B}\,\eta(\mathbf{r}_\text{B}) \tens{G}^{(0)}\left(
\mathbf{r},\mathbf{r}_\text{B} \right)
 \cdot \bm{\alpha}_\text{B}\left(0 \right) \cdot \tens{G}^{(0)}\left( \mathbf{r}_\text{B},\mathbf{r}'\right) + ...
\end{equation}
and subsequently in terms of the electrostatic Green's function:
\begin{equation}
g^{(1)}\left( \mathbf{r},\mathbf{r}'  \right)=-
\frac{1}{\epsilon_0}\int \text{d}^3r_\text{B}\,\eta(\mathbf{r}_\text{B}) \nabla_\text{B} g^{(0)}\left(
\mathbf{r},\mathbf{r}_\text{B} \right)
 \cdot \bm{\alpha}_\text{B}\left(0 \right) \cdot \nabla_\text{B}g^{(0)}\left( \mathbf{r}_\text{B},\mathbf{r}'\right) + ...
\end{equation}
We substitute this expansion into the single-particle Coulomb interaction, Eq.~\eqref{egreenA}, and consider the free-space Green function $g^{\left( 0 \right)}\left( \mathbf{r},\mathbf{r}' \right) = 1/(4\pi \left| \mathbf{r} - \mathbf{r}' \right|)$ as the background to be perturbed. We find:
\begin{equation}
U\left( \mathbf{r}_\text{A} \right) =
\int \text{d}^3r_\text{B}\,\eta(\mathbf{r}_\text{B})
U_{q-\alpha}\left( \mathbf{r}_\text{A},\mathbf{r}_\text{B} \right)
\end{equation}
where $U_{q-\alpha}\left( \mathbf{r}_\text{A},\mathbf{r}_\text{B} \right)$ is the electrostatic interaction between a charged particle and a polarizable molecule:
\begin{equation}\label{UPol}
U_{q - \alpha }\left( \mathbf{r}_\text{A},\mathbf{r}_\text{B} \right) =  - \frac{q_\text{A}^2}{32 \pi ^2\epsilon _0^2}\frac{\mathbf{r} \cdot \bm{\alpha} _\text{B}\left( 0 \right) \cdot \mathbf{r}}{r^6}
\end{equation}
and $\textbf{r}=\textbf{r}_\text{A}-\textbf{r}_\text{B}$.
 This expression is well-known in the literature for the special case of isotropic particles \cite{erbil}, in which case one has;
\begin{equation}
U_{q - \alpha }^{\mathrm{is}}\left( \mathbf{r}_\text{A},\mathbf{r}_\text{B} \right)=   - \frac{q_\text{A}^2  \alpha _\text{B}\left( 0 \right)}{32 \pi ^2\epsilon _0^2r^4}
\end{equation}
Our result \eqref{UPol} generalises this to anisotropically polarized particles, as well as being applicable to any geometry.

\section{Conclusions and outlook}
\label{Sec5}

In this article we have developed a systematic and unified description of Coulomb interactions of charges in non-trivial environments.
The presence of the environment is included via the classical Green's tensor, or also in a simpler fashion in terms of the Green's function. \\

Our approach can be applied to non-trivial geometries where it is not possible or practical to find the suitable image charges, we have demonstrated this via the example of the plate with a hole. Using the same framework, we have shown examples where the environment significantly changes the interaction, for example  by exponentially suppressing it in the dielectric cavity geometry. We have quantified how the Coulomb interaction can be significantly tuned by changing the geometric and dielectric parameters of the environment, many more cases of which could be investigated, all within the formalism presented here.\\
 
Thus the outlook from this work is to apply the formalism to important practical examples. For example, we have considered only neutral environments here, while without much extra complication one could consider environments which carry a net charge, like for example ionic solutions. We have also considered only stationary charges, in which case there is no real complication associated with the instantaneous nature of the Coulomb interaction. A time-dependent model could be developed to include retardation effects thereby satisfying causality requirements and opening up our work to the study of moving charges. \\

\appendix
\section{Green's function of a cavity}\label{Appendix}

We consider the cavity configuration in Figure (\ref{fig4}). The dielectric constant is:
\begin{equation} \epsilon \left( z,\rho \right)=\begin{cases}
  \epsilon_1,& \text{if } z<-d/2\\
     \epsilon_2,& \text{if } -d/2<z<d/2\\
    \epsilon_3,& \text{if } z>d/2 
\end{cases}\end{equation}
where $\rho$ is the radial component of a cylindrical co-ordinate system. A point charge is placed at $z=z_0$ in the central region, which induces surface charges at the interfaces. The corresponding source-term of the Green function is:
\begin{equation}
       \frac{1}{4\pi \epsilon _2\left| \mathbf{r} - \mathbf{r}_0 \right|} = \frac{1}{4\pi \epsilon _2}\int\limits_0^\infty  \text{d}k J_0 \left( k\rho  \right) \text{e}^{ - k\left| z - z_0 \right|}
\end{equation}
where $J_n$ is the $n$th Bessel function of the first kind.
In the central region the Green's function consists in a superposition of rising and falling exponentials, as well a term stemming from the point source:
\begin{equation}
g_2=\frac{1}{4\pi \epsilon _2}\int\limits_0^\infty  \text{d}k J_0 \left( k\rho  \right) \Big(a(k) \left( k \right)\text{e}^{ - k(z - z_0) }
+b  (k) \text{e}^{  k(z - z_0) }+\text{e}^{ - k\left| z - z_0 \right|}\Big)
\end{equation}
where $a(k)$ and $b(k)$ are coefficients to be found. In the left region we have only rising exponential since the Green's function must vanish for $z \to -\infty$:
\begin{equation}g_1=\frac{1}{4\pi \epsilon _2}\int\limits_0^\infty  \text{d}k J_0 \left( k\rho  \right)c(k) \text{e}^{  k(z - z_0) }
\end{equation}
while in the right region:
\begin{equation}g_3=\frac{1}{4\pi \epsilon _2}\int\limits_0^\infty  \text{d}k J_0 \left( k\rho  \right)d(k)\text{e}^{  -k(z - z_0) }
\end{equation}
where $c(k)$ and $d(k)$ are coefficients not yet determined. In order to find the four unknown coefficients $a(k),b(k),c(k),d(k)$ we impose the condition from Maxwell's equations that the Green's function and the normal component of the displacement vector $\mathbf{D}$ are continuous across the interface between the media:
\begin{align}\nonumber
g_1 \big|_{z=-d/2}=&g_2 \big|_{z=-d/2}\\\nonumber
\epsilon_1 \partial g_1/\partial z \big|_{z=-d/2}=&\epsilon_2 \partial g_2/\partial z \big|_{z=-d/2} \\\nonumber
g_2 \big|_{z=d/2}=&g_3 \big|_{z=d/2}\\
\epsilon_2 \partial g_2/\partial z \big|_{z=d/2}=&\epsilon_3 \partial g_3/\partial z \big|_{z=d/2}
\end{align}
Solving the resulting system of equations allows one to eliminate the four unknowns, the result for the Green's function in the central region is, for $z \geqslant z_0$:
\begin{equation}
g_2 =\frac{1}{4\pi \epsilon _2}\int\limits_0^\infty  \text{d}k \text{e}^{ - k\left( z - z_0 \right)}
\frac{\left( 1 - \text{e}^{ - k\left( d + 2z_0 \right)}R_1 \right)\left( 1 - \text{e}^{ - k\left( d - 2z \right)}R_1 \right)}{\left( 1 - \text{e}^{ - 2kd}R_1R_3 \right)}J_0\left( k\rho  \right)
\end{equation}
where $R_1$ and $R_3$ are the reflection coefficients for the left and right media, as shown in Eq.~(\ref{R1r3}).

\ack
The authors thank Gabriel Barton and Carsten Henkel for very useful discussions, and acknowledge DFG (grants BU 1803/3-1 and GRK 2079/1). SYB additionally acknowledges support from the Freiburg Institute for Advanced Studies (FRIAS).

%
%
%
%
%
%
%
%
%
%
%
%
%
%


\end{document}